\documentclass[11pt]{article}

\usepackage{epsfig}

\usepackage{graphicx}

\usepackage{amsfonts}
\usepackage{amssymb}
\usepackage{amsthm}
\usepackage{amsmath}

\usepackage{array}

\newcommand{\ba}{\begin{eqnarray}}
\newcommand{\ea}{\end{eqnarray}}
\newcommand{\be}{\begin{equation}}
\newcommand{\ee}{\end{equation}}
\newcommand{\eps}{\epsilon}
\newcommand{\ga}{\alpha}

\newcommand{\ra}{\rightarrow}

\newcommand{\beq}{\begin{equation}}
\newcommand{\eeq}{\end{equation}}
\newcommand{\bea}{\begin{eqnarray}}
\newcommand{\eea}{\end{eqnarray}}

\begin{document}

\thispagestyle{empty}
\vspace*{1cm}
\begin{center}
{\Large{\bf
Towards a realistic Standard
Model from D-brane configurations}}

\vspace*{1cm}
{\large
G.K. Leontaris($^1$), N.D. Tracas($^2$), N.D. Vlachos($^3$), O. Korakianitis($^2$)}

\vspace{0.5cm}
($^1$)Theoretical Physics Division, Ioannina University\\
GR 451 10 Ioannina, Greece\\
($^2$)Physics Department, National Technical University\\
GR 157 73Athens, Greece\\
($^3$)Dept. of Theoretical Physics, Aristotle University of Thessaloniki\\
GR 541 24 Thessaloniki, Greece
\end{center}

\vspace*{2cm}
\begin{abstract}
Effective low energy models arising in the context of
D-brane configurations with Standard Model (SM) gauge symmetry
extended by several gauged abelian factors are discussed. The models
are classified according to their hypercharge embeddings consistent
with the SM spectrum hypercharge assignment. Particular cases are
analyzed according to their perspectives and viability as low energy
effective field theory candidates. The resulting string scale is
determined by means of a two-loop renormalization group calculation.
Their implications in Yukawa couplings, neutrinos and flavor
changing processes are also presented.
\end{abstract}

\vfill
\begin{flushleft}
July 2007
\end{flushleft}
\newpage

\section{Introduction}

The revelation of the higher dimensional
objects~\cite{Polchinski:1995mt}, called D-branes, has
revived the interest on model building in the context of string
theory.  As a consequence, during the last decade, numerous painstaking
efforts  have been devoted to the study of possible D-brane
realizations of the Standard Model and the higher gauge symmetries
containing it\footnote{For a comprehensive and pedagogical
introduction see~\cite{Blumenhagen:2005mu,Blumenhagen:2006ci}.}.

Model building in string theory has shown that  there is no a priori
obvious recipe to obtain the SM from the first principles of the
theory. In recent attempts, various groups
\cite{Blumenhagen:2006ci}-\cite{Kiritsis:2003mc} started mainly a
bottom-up approach delving into the string vacua, aiming to systematically
classify all possible D-brane configurations, seeking an acceptable
effective low energy theory which reproduces the success of the
Standard Model. It is anticipated that, such a construction would
determine  the arbitrary parameters of the Standard Model, while new
phenomena could be predicted and eventually  tested in future
experiments.

In the present work, we elaborate on low energy implications of a particular class
of D-brane models~\cite{Antoniadis:2001np}-\cite{Kiritsis:2003mc}
with Standard Model gauge symmetry and Split Supersymmetric
spectrum~\cite{Arkani-Hamed:2004fb}\footnote{%
For another approach in partly supersymmetric spectrum see ref.\cite{Gherghetta:2003wm}}.
The implementation of Split
Supersymmetry in the D-brane constructions under consideration, is
justified for the following two reasons: First, it was shown that
the realization of Split Susy is a viable possibility in certain
D-brane constructions~\cite{Antoniadis:2004dt}. Second, intermediate
and high string scale D-brane models  previously abandoned
because of phenomenological drawbacks related to hierarchy
problems, rapid proton decay etc, in the context of Split
Supersymmetry offer fascinating new possibilities since  there
exist now convincing arguments concerning the hierarchy issues. Besides,
the renormalization group flow of the gauge and Yukawa couplings as well
as low energy measurable physical quantities dependent on them, change
substantially. In view of these interesting novelties, in a previous
work~\cite{Gioutsos:2005uw},  a classification of  the various D-brane
derived models with Split Supersymmetric spectrum and Standard Model
gauge symmetry extended by $U(1)$ factors was attempted. All possible
configurations with $P=1,2,3$ abelian branes were considered and  the different
hypercharge embeddings compatible with the SM particle spectrum were found. In
all viable cases,  a one-loop RG analysis was used to calculate
the string scale $M_S$, while  the fermion mass relations of
the third generation, the gaugino masses and the lifetime of the
gluino were examined. Here, we will pursue a further investigation, addressing
more phenomenological issues and deriving possible constraints on
the Split Supersymmetry breaking scale $\tilde{m}$ and other so far
undetermined parameters.  We will extend our previous
analysis, and  work out in detail the predictions for the string scale
using renormalization group equations at two--loop order. We note
however, that the determination of the string scale is more
intricate than in ordinary Grand Unified models. The reason is that
in D-brane constructions, gauge coupling unification at the string
scale does not occur since the volume of the internal space is
involved between gauge and string couplings; thus, the actual values
of the SM gauge coupling constants may differ at $M_S$. This
arbitrariness looks rather daunting, however, certain internal
volume relations could allow for partial unification.
 In order to reduce the number of free parameters and obtain
definite predictions, we mainly concentrate on cases where certain
relations are assumed to connect the gauge couplings at the string scale.
Particular attention is also given  to models
where the non-abelian gauge couplings have a common value at $M_S$.
Next, in each case of the models under consideration, we determine
the range of the Split Supersymmetric scale which is compatible with
the chosen gauge coupling conditions. We further discuss the Yukawa
potential and other low energy predictions which are crucial for
the viability of the models. In particular, we examine the
conditions under which  exotic processes like the lepton flavor violating
decay $\mu\ra e\gamma$ can be detected in future experiments, we
discuss suggested mechanisms predicting a sufficiently heavy mass for
the right-handed neutrino and check the existence of exotic matter
like leptoquarks, whose appearance is also possible in the low
energy spectrum of these models.

The paper is organized as follows: In the next section, we summarize
the salient features of the D-brane configurations and determine
the hypercharge embeddings leading to Standard Model particle
spectrum. In section 3 we present in some detail selected cases of  low energy
effective models, while in section 4
we calculate the string scale and explore its correlation with the Split
Supersymmetric scale by means of a two-loop renormalization group analysis.
In section 5 we analyze the effects of  exotic states.
In section 6 we discuss the observability conditions for
rare flavor violation like $\mu\ra e\gamma$.
Finally, in section 7 we present our conclusions.

\section{ D-brane configurations  and Split Supersymmetry}

 The embedding of the Standard Model in a D-brane configuration as well as
 some implications in low energy phenomenology and the magnitude of the
 string scale have been
explored in  several works~ \cite{Antoniadis:2001np,
Antoniadis:2002qm,Gioutsos:2005uw,Anastasopoulos:2006da,Gioutsos:2006fv}.
The same problem in the context of intersecting D-branes, has also been
extensively discussed~\cite{Blumenhagen:2000wh,Aldazabal:2000cn}.

 The resulting field theory model involves the SM non-abelian gauge symmetry
  extended by several $U(1)$ factors, a linear combination of which
  defines the hypercharge.\footnote{see reviews
\cite{Kiritsis:2003mc,Blumenhagen:2006ci}.} A systematic  bottom-up
investigation  of all possible configurations for the SM gauge
symmetry with two abelian branes was presented
in~\cite{Anastasopoulos:2006da}. Here, as in \cite{Gioutsos:2006fv},
we assume the existence of at most three extra $U(1)$ abelian
branes, thus, the full gauge group is
\ba%
G={U(3)}_C\times{U(2)}_L\times{U(1)}^P,\; P=1,2,3 \, \cdot \label{ggg}
\ea
Since $U(n)\sim {SU(n)}\times{U(1)}$, the final symmetry is
\ba%
G={SU(3)}_C\times{SU(2)}_L\times{U(1)}^{P+2},\; P=1,2,3 \label{ggg1}
\ea
while the SM fermions and Higgs fields carry additional quantum
numbers under the extra $U(1)$'s.

Given that strings attached to various D-brane stacks represent the SM
matter fields, the hypercharge generator is in general, a linear
combination of all possible $U(1)$ factors.  If we define the
anomaly free linear combination of these $U(1)$'s to be
$$Y=k_3Q_3+k_2Q_2+\sum_{i=1}^3k_i'Q_i'$$ the most general
hypercharge gauge coupling condition can be written as
\ba%
\frac{1}{g_Y^2}=\frac{6 k_3^2}{g_3^2}+\frac{4
k_2^2}{g_2^2}+2\sum_{i=1}^P
\frac{{k'_i}^2}{{g'_i}^2}\ \cdot \label{gY}%
\ea%
For a given hypercharge embedding, the $k_i'$'s can be determined and
equation (\ref{gY}) relates the weak angle
$\sin^2\theta_W=({1+k_Y})^{-1}$ ( $k_Y =
\frac{\alpha_2}{\alpha_Y}$,  $\alpha_i \equiv g_i^2/4\pi$),
with the gauge coupling ratios at the string scale $M_S$.  Since in
D-brane scenarios the gauge couplings do not necessarily attain a
common value at the string (brane) scale $M_S$, the ratios
$\alpha_2/\alpha_3$, $\alpha_2/\alpha_i'$ differ from unity there.
In a previous analysis, various relations between the gauge couplings
were considered and a systematic investigation of the magnitude
of the string scale and the split SUSY breaking mass scale was
presented~\cite{Gioutsos:2006fv}. Here, in order to reduce the
arbitrary parameters and come up with definite predictions, we assume
the existence of relations between the gauge couplings at $M_S$, while we
pay particular attention to the interesting case $\alpha_2=\alpha_3$
at $M_S$. If we  accept that $m\le P$ $U(1)$ branes are aligned with
the $U(3)$ stack and the remaining $P-m$ abelian branes are aligned
with the $U(2)$ set, $k_Y$ becomes
\ba
k_Y& \equiv& \frac{\alpha_2}{\alpha_Y}
   \;=\;n_1\,\xi+n_2\label{kY}
\ea
where
\ba
n_1 ={6 k_3^2} +2\sum_{i=1}^{m} {k_i'}^2 ,&&
n_2=4k_2^2+2\sum_{i=m+1}^{P}{k_i'}^2 ,\label{xyz}
\ea
 and $\xi$ is the ratio of the
non-abelian gauge couplings ${\alpha_2}/{\alpha_3}$.

\begin{figure}[!t]
\centering
\includegraphics[width=0.3\textwidth]{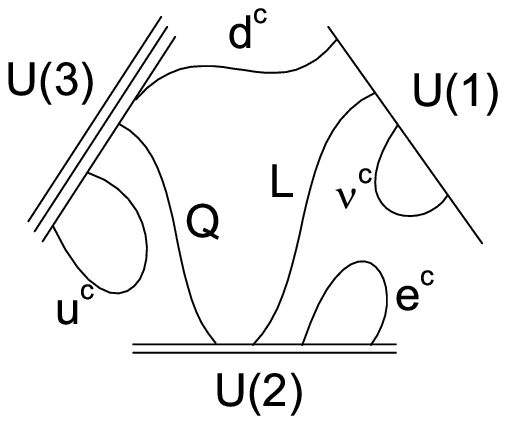}
\includegraphics[width=0.3\textwidth]{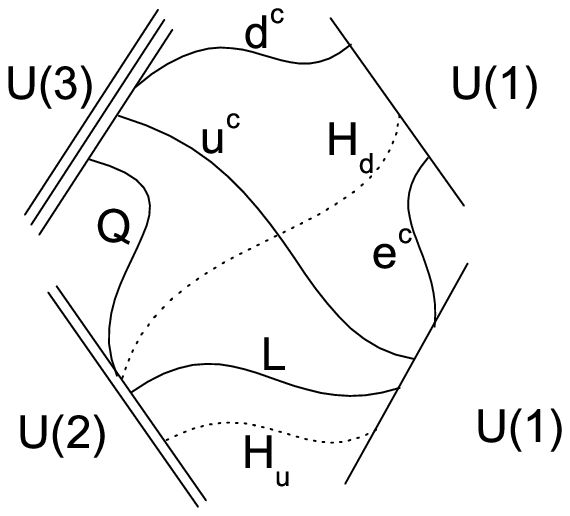}
\includegraphics[width=0.3\textwidth]{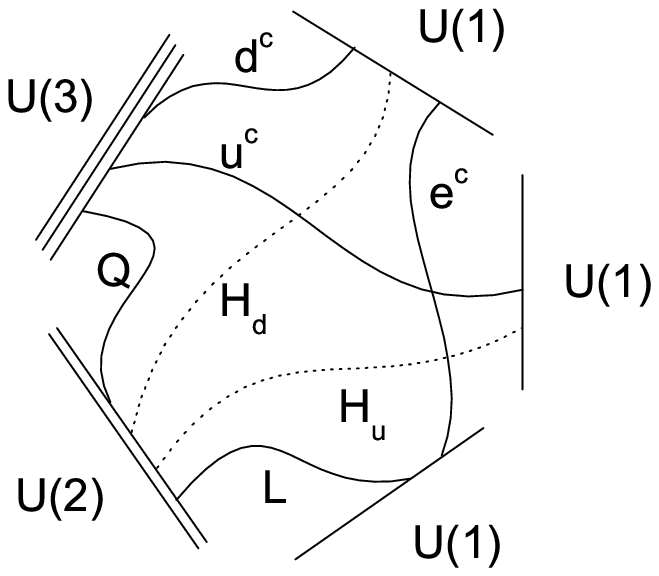}
\caption{\label{f23}Selected Standard Model configurations with one,
two and three
 abelian branes $P=1,2,3$.}
\end{figure}

To obtain the fermion and Higgs spectrum of a given brane
configuration, we note that each state corresponds to an open string
stretched between pairs of brane stacks or to a string with both
ends attached to the same brane stack. Taking all possible
arrangements of the SM particle spectrum represented by the various
strings attached between the $U(3)$, $U(2)$ and extra $U(1)_i$ brane
sets, we end up with the admissible brane  configurations. Some
particular arrangements are shown in figure  \ref{f23}. For each
particular configuration, the coefficients $k_{2,3}, k_i'$ are
determined by the requirement that the  SM particle spectrum acquires
the correct hypercharge.

To proceed further, we recall here the results obtained for brane
setups that include up to three abelian branes. The admissible solutions for the
 $U(3)\times U(2)\times U(1)$ case setup have been explored in
~\cite{Antoniadis:2004dt} and the $k_i$ distinctive solutions are
presented in the first two lines of Table~\ref{t1}. For the
$U(3)\times U(2)\times U(1)^2$
configuration~\cite{Antoniadis:2001np,Antoniadis:2002qm,Anastasopoulos:2006da},
we assign the quantum numbers of the SM particles
$Q(3,2;+1,\eps_1,0,0)$, $d^c(\bar 3,1;-1,0,\eps_2,0)$, $u^c(\bar
3,1;-1,0,0,-\eps_3)$, $L(1,2;0,\eps_4,0,\eps_5)$ and
$e^c(1,1;0,0,\eps_6,\eps_7)$, where $\eps_i=\pm 1$. A similar
assignment can be written for the $P=3$ case too. Solving the
corresponding hypercharge assignment
equations~\cite{Gioutsos:2005uw,Gioutsos:2006fv}, we find that the
hypercharge can be expressed in terms of $k_2'=x$ which remains
undetermined, as
\ba
Y&=& (2/3-x) \,Q_3+( 1/2- x)\, Q_2+(1-x)\, Q_1'+x\,
Q_2'+\delta_{N3}\,x\,Q_3'\ \cdot\label{N23}
\ea
Choosing appropriate values for $x$ so that the corresponding boson
remains massless at $M_S$, the simplest solutions for the three
different brane configurations are shown in Table~\ref{t1}. For
future use, the values of the coefficients $n_1,n_2$ appearing in
(\ref{xyz}) for all possible $U(3)$ and/or $U(2)$ alignments of the
$U(1)$ branes are also included in the last column of the same Table.

\begin{table}[!t]
\centering
\renewcommand{\arraystretch}{1.2}
\begin{tabular}{|c|l|c|c|c|c|c||l|}
\hline
$P$ & & $|k_3|$ & $|k_2|$ & $|k_1'|$ & $|k_2'|$ & $|k_3'|$&$(n_1,n_2)$\\
\hline
    & $a_1$ & $\frac 13$ & $\frac 12$ & $0$        & $-$ & $-$&$(\frac 23,1)$\\
\raisebox{1.5ex}[0pt]{$1$} & $b_1$ & $\frac 16$ & $0$        & $\frac 12$ & $-$ & $-$&$(\frac 23,0);(\frac 16,\frac 12)$\\
\hline
    & $a_2$ & $\frac 16$ & $0$        & $\frac 12$ & $\frac 12$ & $-$&$(\frac 16,1);(\frac 76,0);(\frac 23,\frac 12)$\\
$2$ & $b_2$ & $\frac 23$ & $\frac 12$ & $1$        & $0$        &
$-$ &$(\frac 83,3);(\frac{14}3,1)$\\
    & $c_2$ & $\frac 13$ & $\frac 12$ & $0$        & $1$        & $-$&$(\frac 23,3);(\frac{8}3,1)$\\
\hline
    & $a_3$ & $\frac 16$ & $0$        & $\frac 12$ & $\frac 12$  & $\frac 12$&$(\frac 16,\frac 32);(\frac 23,1);(\frac 53,0);(\frac 76,\frac 12)$\\
$3$ & $b_3$ & $\frac 13$ & $\frac 12$ & $0$        & $1$         & $1$&$(\frac 23,5);(\frac{14}3,1);(\frac 83,3)$\\
    & $c_3$ & $\frac 23$ & $\frac 12$ & $1$        & $0$         & $0$&$(\frac 83,3);(\frac{14}3,1)$\\
\hline
\end{tabular}
\caption{\label{t1}Simplest hypercharge embeddings for the $P=1,2,3$
brane configurations. The last column shows the $(n_1,n_2)$-values
of (\ref{xyz}) for the various possible alignments of the
$U(1)$-branes with respect to $U(3)$ and $U(2)$ brane-stack
orientations.}
\end{table}

\section{Low energy effective models of particular embeddings }

In the previous section we saw that the SM spectrum can be
successfully accommodated in  D-brane setups with gauge symmetry of
the form $U(3)\times U(2)\times U(1)^P$, $P= 1,2,3$, and made a complete classification
of models for the various hypercharge embeddings which
imply a realistic particle content. In this section, we will
present the particular embeddings in more detail, and work out
the Yukawa couplings and other phenomenologically interesting features.

In a previous work~\cite{Gioutsos:2006fv}, we presented an analysis
dealing with the implications of Split Supersymmetry on the string scale,
first considering  models that arise in parallel brane scenarios
where the $U(1)$ branes are superposed with the $U(2)$ or $U(3)$
brane stacks. Varying the Split Susy scale in a wide range, we
examined the evolution of the gauge couplings
and found three distinct classes of models with respect to the
string scale. The Split  Supersymmetry breaking scale $\tilde m$, can
in principle  be much larger than the electroweak scale and, as a
consequence, squarks and sleptons can obtain large masses of  order
$\tilde m$, while the corresponding fermionic degrees as well as
gauginos and higgsinos, remain light. This splitting of the spectrum
is made possible only when the dominant source of supersymmetry breaking
preserves an R-symmetry which protects fermionic degrees from obtaining
masses at the scale $\tilde m$.

 Assuming that above  $\tilde m$  only the MSSM spectrum exists, while below
  $\tilde m$ we only have SM fermions, gauginos, higgsinos and one linear combination
of the scalar Higgs doublets, the one-loop RGE  expression for the
string scale is given by
\ba%
M_S&=& M_Z  \, \left(\frac{\tilde
m}{M_Z}\right)^{a}e^{b}\label{stringscale}
\ea%
where, the parameters $a,b$ depend on the beta function coefficients,
the values of the gauge couplings at $M_Z$ and the model dependent
constants  $n_1,n_2$ given in (\ref{xyz}). After substituting the beta functions,
these constants are given by
\ba
a&=& \frac{21-12n_1-13n_2}{6(11+3n_1-n_2)}\nonumber
\\
b&=& \frac{2\pi}{11+3n_1-n_2} \left( \frac{1}{\alpha_Y} -
\frac{n_2}{\alpha_2}- \frac{n_1}{\alpha_3} \right) \nonumber \cdot
\ea%

 The above one-loop formula is sufficient to produce some qualitative results.
 Thus, one class of models that arises in the configurations
with $P=1$ and $P=3$ abelian branes, predicts a string scale of the
order of the SUSY GUT scale $M_S\sim 10^{16}$ GeV; interestingly enough,
these models also imply that the non-abelian gauge couplings unify
($\alpha_2=\alpha_3$) at $M_S$.

 In the case of the configuration with only one abelian brane, the up or
down right handed quarks arise when both  endpoints of an open
string are attached to the color stack. In this case, the terms
$Qu^c$ or $Qd^c$ carry a non-zero $U(1)_C$ charge while a typical SM
Higgs field is not charged under $U(1)_C$. Therefore, the
corresponding tree-level Yukawa term is not allowed and the quark
fields remain massless.  It is worth noting that, as  shown by  one-loop
renormalization group analysis \cite{Gioutsos:2006fv}, $\tilde{m}$, at least in its
minimal SM content, is fixed at a relatively small scale $\sim 6$ TeV.
The corresponding --with respect to the $M_S\sim 10^{16}$ GeV
prediction-- $P=3$ case, is free from these shortcomings,
consequently, in what follows, we are going to further elaborate on some
interesting low energy implications of this setup.

Another category of brane configurations was found in a particular
--with respect to the $U(1)$ alignments-- case of the $P=2$ abelian
branes. This model corresponds to a specific $U(1)$ brane
orientation and predicts an intermediate string scale $M_S\sim
10^6-10^7$ GeV. Finally, two more cases in the $P=2$ and $P=3$ abelian brane
scenarios result in a low $M_S$, in the TeV range. In the following, we shall give a detailed
description of
 two promising cases with three and two abelian
branes respectively.

\subsection{A representative case  with $U(3)\times U(2)\times U(1)^3$ gauge symmetry}
This model possesses interesting characteristics and enough freedom to produce reliable
phenomenology. The hypercharge assignment
 corresponds to the solution  $a_3$ of Table 1. Depending on
the orientations of the $U(1)$ branes which are taken to be either
parallel to $U(3)$ or to $U(2)$ stack, we obtain four distinct
cases. In particular, if we align the two $U(1)$ branes with the
$U(2)$ stack, we get $k_Y=\frac 16\xi+\frac 32$, and one obtains  non-abelian
gauge unification $\xi=\frac{\alpha_2}{\alpha_3}=1$ at a scale
$M_S\approx 10^{16}$ GeV.   It is worth noting that this particular
$U(1)$ brane alignment results to a $U(1)$ normalization constant
$k_Y=\frac 53$ and $\sin^2\theta_w(M_S)=\frac 38$. These are
undeniably interesting attributes reminiscent of the successful GUTs,
while, in addition, a sufficiently
large mass for the right-handed neutrino arises to realize the
see-saw mechanism. Note also, that all other possible $U(1)$
alignments of the $P=3$ setup lead to a similar unification scale,
but with a weak dependence on the $\tilde m$ scale. Demanding the
existence of  appropriate Yukawa couplings, the various signs are
fixed and this configuration leads to the charge assignments
presented in Table~\ref{tN3}. The field $\nu^c$ in particular
included in the spectrum, is a generic type of singlet which may
arise from a string with endpoints attached on two different $U(1)$ branes,
thus, the possible values of $s_{i=1,2,3}$ are equal to $0,\pm 1$. Once the two
branes where the corresponding string is attached are specified, the
values of $s_i$, can be chosen so that $\nu^c$ can be identified with a RH neutrino.
The hypercharge operator is defined (see
Table 1) as follows:
\ba
Y&=& \frac 16\,Q_c+\frac 12\,\left(Q_1+Q_1'+Q''_1\right)\ \cdot
\label{HQa3}
\ea
Taking into account the charge assignment of the SM states presented
in Table~\ref{tN3}, we can derive the allowed tree-level Yukawa
couplings for the charged fermions which are
\ba
\lambda_u\,Q\,u^c\,H_u+\lambda_d\,Q\,d^c\,H_d+\lambda_l\,L\,e^c\,H_d \, \cdot
\label{pot}
\ea
The potential mass terms (\ref{pot}) do not discriminate between generations
therefore some other mechanism has to be invented in order to generate flavor
mixing.
This can be achieved in the case of intersecting branes, where
quarks and leptons as well as Higgs fields appear at the
intersections and are located at different positions in the compact
extra dimensions~\cite{Blumenhagen:2000wh},\cite{Cremades:2003qj}.
The six dimensional compact space is usually taken to be a
six-dimensional factorizable torus $T^6=\prod_{i=1}3 T^2_i$ while
strings representing the matter fields are wrapped along the two
1-cycles of each of the three torii. The number of fermion
generations is related to the two distinct numbers of brane
wrappings around the two circles of the three torii. For example,
for a string with endpoints attached on two stacks $a, b$,
(corresponding to a $(N_a,\bar N_b)$ representation)  and  wrapping numbers
$(n_a^i,m_a^i)$,$(n_b^i,m_b^i)$, the number of intersections
$I_{ab}=\prod_{i=1}3(n_a^im_b^i-n_b^im_a^i)$ equals the number of
chiral fermions at the intersection. In this scenario, the trilinear
flavor mixing Yukawa couplings of the form
$\lambda^{ijk}f_i^cf_jh_k$ arise from a string world-sheet
stretching between the three relevant brane stacks, while the
coupling strengths are of the order $\lambda^{ijk}\sim
e^{-A_{ijk}}$, where $A_{ijk}$ is the  triangular area generated by
the three vertices related to the fermions $f_i^c,f_j$ and the Higgs
field $h_k$.
We note,
however,
that in the present model one might exploit
the fact that the three $U(1)$ charges $Q_1,Q_1'$  and $Q_1''$
appear symmetrically in the hypercharge definition. This allows for the
possibility of having open strings with one end attached to a certain
non-abelian stack and the other endpoint attached to a different $U(1)$
brane.  In this case, the corresponding SM states have the same
$SU(3)\times SU(2)\times U(1)_Y$ quantum numbers, although they are
differently charged under the extra $U(1)$'s. The latter could act
as a family symmetry distinguishing between the various -`flavor
dependent'- Yukawa terms. As an example, assume that in addition to
the string representing the $u^c$ of Table (\ref{tN3}), we also add
a string with one end attached to the $U(3)$ stack and the other end to
the first $U(1)$ brane with quantum numbers $(\bar
3,1;-1,-1,0,0,0)$. This could also be interpreted  as a right-handed
up-quark field (denoted here as $u'^c$)  belonging to a different
family, however another tree level term $Qu'^c H_u$ is prevented by the $U(1)$ symmetry.
 A mass term  for this $u'^c$ could be possible in
the presence of a neutral scalar singlet $(1,1;-1,1,0)$ (represented
by a string with ends on the appropriate $U(1)$ branes) so that a
hierarchically suppressed mass term of the form
$Qu'^cH_u\frac{\langle\phi\rangle}{M_S}$ could arise. Similar terms
can also be generated for the lepton fields.
\begin{table}[!t]
\centering
\renewcommand{\arraystretch}{1.2}
\begin{tabular}{|c|c|ccccc|}
\hline
 & $SU(3)\times SU(2)$& $Q_3$ & $Q_2$ & $Q_1'$ & $Q_2'$ & $Q_3'$\\
\hline
    & $Q\,(3,2)$ & $1$ & $\eps_1$ & $0$        & $0$ & $0$\\
& $u^c\,(\bar 3,1)$ & $-1$ & $0$        & $0$ & $-1$ & $0$\\
   Fermions  & $d^c\,(\bar 3,1)$ & $-1$ & $0$        & $1$ & $0$ & $0$\\
& $L\,(1,2)$ & $0$ & $\eps_1$ & $0$        & $0$        & $-1$\\
    & $e^c\,(1,1)$ & $0$ & $0$ & $1$        & $0$        & $1$\\
     & $\nu^c\,(1,1)$ & $0$ & $0$ & $s_1$        & $s_2$         & $s_3$\\
\hline
    & $H_u\,(1,2)$ & $0$ & $-\eps_1$        & $0$ & $1$  & $0$\\
Bosons & $H_d\,(1,2)$ & $0$ & $-\eps_1$ & $-1$       & $0$         & $0$ \\
\hline
\end{tabular}
\caption{\label{tN3}The quantum numbers of the SM matter fields for
the $N=3$ brane configuration. The particular choice of signs
 are fixed by hypercharge and Yukawa couplings
constraints.}
\end{table}

Returning now to the minimal version of the present model, it can be
easily checked that the Lepton number is identified with the ${\cal
L}=-Q_3'$ charge of the above states, so that the lepton doublet $L$
and the singlet $e^c$ carry lepton numbers $+1$ and $-1$
respectively, while all other states are neutral under $Q_3'$.
Taking into account this definition of the Lepton number, we
deduce that there are only two possibilities to accommodate the
right handed neutrino. If we choose $s_1=0, s_2=-1,s_3=+1$, we get
 a $\nu^c$ state with $U(1)$ charges $(0,0,0,-1,1)$ and zero
hypercharge. Then a Yukawa coupling of the form
\ba
\lambda_l\,L\,H_u\,\nu^c
\ea
is compatible with all $U(1)$ symmetries, providing
the neutrino with Dirac mass. This Dirac mass term is naturally of the same order of
magnitude as the corresponding mass terms for charged lepton fields. Its suppression down to the
present experimental limits, is achieved via the see-saw
mechanism, therefore a Majorana neutrino mass term
$M_{\nu^c}\,\nu^c\,\nu^c$ with $M_{\nu^c}\sim M_S$ is required.
Unfortunately, a mass scale of this high order in not possible to get in
the perturbative superpotential.  A non-perturbative origin for
$M_{\nu^c}$ and in particular from String Theory instanton effects
was proposed in \cite{Ibanez:2006da}. According to this approach, the
operator
\ba
M_S\,e^{-S_{inst.}}\,\nu^c\,\nu^c\label{npnc}
\ea
which provides a Majorana mass to $\nu^c$ is found to be gauge
invariant, while it violates the $B-L$ symmetry.  The relevant
instantons correspond to D2-branes which, when intersecting with
D6-brane stacks give rise to fermionic zero modes charged under
particular $U(1)$'s. An instanton induced effective interaction is
generated by integrating over the instanton zero modes. Fermion zero
modes charged under the particular $U(1)$ violate the corresponding
$U(1)$ symmetry, therefore it is necessary to insert fields charged under the 4d
symmetry. In the case under consideration, a
D2-brane when intersecting with the two D6 branes where the string
representing $\nu^c$ is attached, gives rise to a Majorana mass term
of the form (\ref{npnc}), thus the see-saw mechanism is operative in
this model.

It has been
observed \cite{Anastasopoulos:2006da},\cite{Ibanez:2007rs} that in
addition to the SM particle spectrum discussed above, in several
D-brane constructions, states which carry both lepton and quark quantum
numbers are unavoidable. Indeed,  fields of the type $D,U=(3/\bar 3,1)_{(\pm
1,0,0,0,1)}$ and their complex conjugates obtained by strings
attached on $U(3)$ and the corresponding $U(1)$ stacks, are also
possible. These representations have the quantum numbers of
leptoquarks, i.e., they are color triplets and carry lepton
number ${\cal L}= \pm 1$. Fortunately,  couplings of the form $\bar
D\,Q\,L$ and $ D\,u^c\,e^c$ that might lead to baryon instability
 are not allowed because of $U(1)$ symmetries.
Nevertheless, these states do contribute to the beta function
coefficients and thus, can have a significant impact on the
determination of the string and Split Susy scales. In the
following sections, a more general analysis that will also
include leptoquarks will be presented.

\subsection{A  case  with $U(3)\times U(2)\times U(1)^2$ gauge symmetry and low $M_S$}

We will now describe a model which admits a low unification scale\footnote{%
For implications of low string scale models see ref.\cite{Barger:2007ay}}. We consider the
case $b_2$ of Table \ref{t1}  where both  $U(1)$ branes are aligned
to the $U(3)$ stack. A rough estimate of the unification scale at
one loop order gives $M_S\sim 10^5\times (m_Z/\tilde m)^{1/3}$,
and it can be checked (\ref{stringscale}) that its highest value
cannot exceed $10^5$ GeV.

The hypercharge assignments of the SM states $Q(3,1;1,-1,0,0)$,
$u^c(\bar 3,1;-1,0,0,\eps_3)$, $d^c(\bar 3,-1,0,1,0)$,
$L(1,2;0,-1,0,\eps_5)$ and $e^c(1,1;0,0,1,\eps_7)$ are consistent
with the hypercharge definition
\ba
Y&=& \frac 23\,Q_3+\frac 12\,Q_2+Q_1' \ \cdot\label{HQb2}
\ea
 The remaining coefficients $\eps_i=\pm 1$ are correlated through the
superpotential terms needed  to generate masses for quarks and
lepton fields; the relevant Yukawa couplings are
\ba
{\cal
W}&=&\lambda_u\,Q\,u^c\,H_u+\lambda_d\,Q\,d^c\,H_d+\lambda_l\,L\,H_d\,e^c
\ea
and imply the relations $ \eps_3=-\eps_9,
 \eps_5=-\eps_7$.
We  may  define the baryon number to be $Q_B=\frac 13\, Q_C$,
whilst, for the given   configuration, the lepton number
can be a combination of the form
\ba
Q_{\cal L}&=& a \left(Q_3+Q_2+Q_1'-\eps_3 \frac{1+a}{a} Q_2'\right)
\ea
where $a$ is a parameter to be specified. Demanding that the fermion
and Higgs fields have the appropriate lepton number one finds that
$a=-\frac{1}{2},\eps_3=+1 $ thus
\ba
Q_{\cal L}&=& -\frac 12 (Q_2+Q_2+Q_1'+  Q_2')\ \cdot
\ea
Finally, taking into account all the constraints, it is found that
the $\eps_{3,5,7,9}$ can be expressed in terms of one parameter only
$ \eps_3=\eps_7=-\eps_5=-\eps_9=+1\nonumber $, while the resulting
charge assignments are shown in Table \ref{t4}.

\begin{table}[!t]
\centering
\renewcommand{\arraystretch}{1.2}
\begin{tabular}{|c|c|cccc|}
\hline
 & $SU(3)\times SU(2)$& $Q_3$ & $Q_2$ & $Q_1'$ & $Q_2'$ \\
\hline
    & $Q\,(3,2)$ & $1$ & $-1$ & $0$        & $0$ \\
& $u^c\,(\bar 3,1)$ & $-1$ & $0$        & $0$ & $1$  \\
   Fermions  & $d^c\,(\bar 3,1)$ & $-1$ & $0$        & $1$ & $0$ \\
& $L\,(1,2)$ & $0$ & $-1$ & $0$        & $-1$        \\
    & $e^c\,(1,1)$ & $0$ & $0$ & $1$        & $1$        \\
\hline
    & $H_u\,(1,2)$ & $0$ & $1$        & $0$ & $-1$  \\
Bosons & $H_d\,(1,2)$ & $0$ & $1$ & $-1$       & $0$         \\
\hline
\end{tabular}
\caption{\label{t4}The quantum numbers of the SM mater fields for
the $N=2$ brane configuration.}
\end{table}

Last, the model should also predict a sufficiently
suppressed neutrino mass. This may be achieved through a see-saw
type mechanism which requires the presence of a right-handed
neutrino with both Dirac and Majonana masses. The existence of a
Dirac mass term
\ba
\lambda_l\,L\,H_u\,\nu^c
\ea
is possible if the RH-neutrino is produced by a string with both
ends attached to the same $U(1)$ (bulk) brane with $U(1)$ charges
$(0,0,0,2)$ which is part of a vector multiplet as described
in~\cite{Antoniadis:2002qm}. We also note that all dangerous Yukawa
terms  of the form
\ba
\lambda_1\,Q\,d^c\,L+\lambda_2\,u^c\,d^c\,d^c+\lambda_3\,L\,L\,e^c
\ea
are forbidden by the $U(1)$ symmetries, unless extra singlets
generate them at higher order. Such terms can also appear through
instanton effects, as it is the case for the  $\nu^c$-mass term.

\section{The two-loop Renormalization Group Analysis}

The bottom up approach in analyzing SM like models derived
from brane setups,  reveals that the number of the additional $U(1)$
branes as well as their
 orientation  with respect to the $U(3)$ and $U(2)$ stacks
have a significant impact on the determination of the string scale.
 Up till now, the analysis was restricted to one-loop order.
 Several issues related to measurements at
 experimentally accessible energies however, require a more refined analysis
 for the various mass scales
 of the theory. In this section, we will proceed to a two-loop
 renormalization group calculation in order to get a more accurate value  for the
 string scale as a function of the Split Susy mass, while, in the next
  section, these new results will be used to determine sensitive quantities
 such as branching ratios for flavor violating processes for which strict experimental bounds exist.

In the present approach, apart from the electroweak scale $M_Z$,
we assume the existence of two additional mass scales,  the string $M_{S}$
and the Split Susy $\tilde{m}$ scales. The string scale is
  defined through the gauge couplings relation
\ba
\frac{1}{\alpha_1}=\frac{n_1}{\alpha_3}+\frac{n_2}{\alpha_2}
\label{atstring}
\ea
subject to the `naturality' condition $\alpha_2\le \alpha_3$ at
$M_S$. This condition ensures that the couplings $\alpha_2,
\alpha_3$ do not meet at a scale lower than the one defined by
(\ref{atstring}). If $\alpha_2= \alpha_3$ is realized prior to the
condition (\ref{atstring}), the scale $M_S$ is then defined at the
point where these two couplings meet.
For a given D-brane configuration, $n_1$ and $n_2$ are
expressed in terms of the particular values of $k_i,k_i'$ in
(\ref{xyz}), and their values are given in the last column of
Table 1.

In order to define $\tilde{m}$,  we assume for
simplicity that all scalars acquire a common mass at a scale between
the string and the electroweak scale.  This scale is
identified with $\tilde{m}$, below which
only gauginos and higgsinos survive from the SUSY spectrum. All viable
models in the context described previously will be considered.
 We will start with models containing
 the  MSSM spectrum, however, our
 analysis will be extended to include exotic
states like leptoquarks, which {\it do} arise in certain D-brane
constructions.

The 2-loop renormalization group equations
for the gauge couplings are given by
\begin{equation}\label{RGE}
\frac{d\ga_i}{dt}=\frac{2b_i\ga_i^2}{4\pi}+
\frac{2\ga_i^2}{(4\pi)^2} \left[ \sum_{j=1}^3 b_{ij}\ga_j -
\sum_{k=u,d,e} \frac{d_i^k}{4\pi}\textrm{Tr}\left(h^{k\dag}
h^k\right) -d^W_i\left(\tilde{\ga}_u+\tilde{\ga}_d\right)-
  d^B_i\left(\tilde{\ga}_u^\prime +\tilde{\ga}_d^\prime \right)
\right]
\end{equation}
where $\tilde{\ga}_u$, $\tilde{\ga}_d$, $\tilde{\ga}_u^\prime$ and
$\tilde{\ga}_d^\prime$ stand for the gaugino couplings appearing in
the Split Susy Lagrangian, $\alpha_i$ ($i=1,2,3$) are the gauge couplings and $h^k$
($k=u,d,e$) are the Yukawa couplings. Below  $\tilde{m}$  the
coefficients of Eq.\ref{RGE} are given by~\cite{Giudice:2004tc}
\begin{equation}\label{coef}
\begin{split}
b_i&=\left(\frac{15}{2},-\frac76,-5\right),\quad b_{ij}=\left(
\begin{array}{ccc}
\frac{104}{9} & 6 & \frac{44}{3}\\
2  & \frac{106}{3}  & 12\\
\frac{11}{6}  & \frac92  & 22
\end{array}
\right)
\\
d^u_i&=\left(\frac{17}{6}, \frac32,2\right),\quad
d^d_i=\left(\frac{5}{6}, \frac32,2\right),\quad
d^e_i=\left(\frac52, \frac12,0\right),\\
d^W_i&=\left(\frac34,\frac{11}{4},0\right),\quad
d^B_i=\left(\frac14,\frac{1}{4},0\right) \, \cdot
\end{split}
\end{equation}
Above $\tilde{m}$  we have the usual MSSM spectrum and the coefficients are:
\begin{equation}\label{Scoef}
\begin{split}
b_i&=\left(11,1,-3\right),\quad b_{ij}=\left(
\begin{array}{ccc}
\frac{199}{9} & 9 & \frac{88}{3}\\
3 & 25  & 24\\
\frac{11}{3}  & 9  & 14
\end{array}
\right)
\\
d^u_i&=\left(\frac{26}{3}, 6,4\right),\quad
d^d_i=\left(\frac{14}{3}, 6,4\right),\quad d^e_i=\left(6,
2,0\right),\quad d^W=0,\quad d^B=0  \, \cdot
\end{split}
\end{equation}
The proper treatment of the two loop RGE's for gauge couplings,
require also the inclusion of the one-loop running for the Yukawa
couplings.  Thus, the Yukawa coupling RGE's below  $\tilde{m}$
 are given  by
\begin{align}\label{Yukawa}
\frac{dh^u}{dt}&=\frac{h^u}{4\pi} \left(-3\sum_{i=1}^3
c_i^u\ga_i+\frac{1}{4\pi}\frac32h^{u\dag}h^u-\frac{1}{4\pi}\frac32h^{d\dag}h^d+\frac{1}{4\pi}T
\right)\\
\frac{dh^d}{dt}&=\frac{h^d}{4\pi} \left(-3\sum_{i=1}^3
c_i^d\ga_i-\frac{1}{4\pi}\frac32h^{u\dag}h^u+\frac{1}{4\pi}\frac32h^{d\dag}h^d+\frac{1}{4\pi}T
\right)\\
\frac{dh^e}{dt}&=\frac{h^e}{4\pi} \left(-3\sum_{i=1}^3
c_i^e\ga_i+\frac{1}{4\pi}\frac32h^{e\dag}h^e+\frac{1}{4\pi}T \right)
\end{align}
where
\begin{equation}
\begin{split}
\frac{1}{4\pi}T&=\frac{1}{4\pi}\textrm{Tr}\left(3h^{u\dag}h^u+3h^{d\dag}h^d+3h^{e\dag}h^e\right)+
\frac32\left(\tilde{\ga}_u+\tilde{\ga}_d\right)+\frac12\left(\tilde{\ga}_u^\prime+\tilde{\ga}_d^\prime\right)\\
c^u_i&=\left(\frac{17}{36},\frac34,\frac83\right),\quad
c^d_i=\left(\frac{5}{36},\frac34,\frac83\right),\quad
c^e_i=\left(\frac34,\frac34,0\right) \, \cdot
\end{split}
\end{equation}

Above  $\tilde{m}$ scale the RGE's are (now $h^k $ is replaced by $\lambda^k$)
\begin{align}
\frac{d\lambda^u}{dt}&=\frac{\lambda^u}{4\pi} \left(-2\sum_{i=1}^3
c_i^u\ga_i+\frac{1}{4\pi}\lambda^{u\dag}\lambda^u+\lambda^{d\prime}\lambda^u+
\frac{1}{4\pi}3\textrm{Tr}(\lambda^{u\dag}\lambda^u)\right)\\
\frac{d\lambda^d}{dt}&=\frac{\lambda^d}{4\pi} \left(-2\sum_{i=1}^3
c_i^d\ga_i+\frac{1}{4\pi}3\lambda^{u\dag}\lambda^u+
\frac{1}{4\pi}\textrm{Tr}(3\lambda^{d\dag}\lambda^d+\lambda^{e\dag}\lambda^e)\right)\\
\frac{d\lambda^e}{dt}&=\frac{\lambda^e}{4\pi} \left(-2\sum_{i=1}^3
c_i^e\ga_i+\frac{1}{4\pi}3\lambda^{e\dag}\lambda^e+
\frac{1}{4\pi}\textrm{Tr}(3\lambda^{d\dag}\lambda^d+\lambda^{e\dag}\lambda^e)\right)
\end{align}
where
\begin{equation}
c^u_i=\left(\frac{13}{18},\frac32,\frac83\right),\quad
c^d_i=\left(\frac{7}{18},\frac32,\frac83\right),\quad
c^e_i=\left(\frac32,\frac32,0\right) \, \cdot
\end{equation}

Finally, the gaugino coupling RGE's are
\begin{align}
\frac{d\tilde{\ga}_u}{dt}&=\frac{2\tilde{\ga}_u}{4\pi} \left(
-3\sum_{i=1}^3 C_i\ga_i
+\frac54\tilde{\ga}_u-\frac12\tilde{\ga}_d+\frac14\tilde{\ga}_u^\prime\right)+
\frac{2}{4\pi}\sqrt{\tilde{\ga}_u\tilde{\ga}_d\tilde{\ga}_u^\prime\tilde{\ga}_d^\prime}+\frac{2\tilde{\ga}_u}{(4\pi)^2}T\\
\frac{d\tilde{\ga}_u^\prime}{dt}&=\frac{2\tilde{\ga}_u^\prime}{4\pi}
\left( -3\sum_{i=1}^3 C_i^\prime\ga_i
+\frac34\tilde{\ga}_u^\prime+\frac32\tilde{\ga}_d^\prime+\frac34\tilde{\ga}_u\right)+
3\frac{2}{4\pi}\sqrt{\tilde{\ga}_u\tilde{\ga}_d\tilde{\ga}_u^\prime\tilde{\ga}_d^\prime}+\frac{2\tilde{\ga}_u^\prime}{(4\pi)^2}T\\
\frac{d\tilde{\ga}_d}{dt}&=\frac{2\tilde{\ga}_d}{4\pi} \left(
-3\sum_{i=1}^3 C_i\ga_i
+\frac54\tilde{\ga}_d-\frac12\tilde{\ga}_u+\frac14\tilde{\ga}_d^\prime\right)+
\frac{2}{4\pi}\sqrt{\tilde{\ga}_u\tilde{\ga}_d\tilde{\ga}_u^\prime\tilde{\ga}_d^\prime}+\frac{2\tilde{\ga}_d}{(4\pi)^2}T\\
\frac{d\tilde{\ga}_d^\prime}{dt}&=\frac{2\tilde{\ga}_d^\prime}{4\pi}
\left( -3\sum_{i=1}^3 C_i^\prime\ga_i
+\frac34\tilde{\ga}_d^\prime+\frac32\tilde{\ga}_u^\prime+\frac34\tilde{\ga}_d\right)+
3\frac{2}{4\pi}\sqrt{\tilde{\ga}_u\tilde{\ga}_d\tilde{\ga}_u^\prime\tilde{\ga}_d^\prime}+\frac{2\tilde{\ga}_d^\prime}{(4\pi)^2}T
\end{align}
where
\[
C_i=\left(\frac14,\frac{11}{4},0\right),\qquad
C^\prime_i=\left(\frac14,\frac34,0\right)\, \cdot
\]

The matching conditions at $\tilde{m}$ are
\begin{equation}\label{matching}
\begin{split}
\tilde{\ga}_u=\ga_2\sin^2\beta, &\qquad  \tilde{\ga}_d=\ga_2\cos^2\beta\\
\tilde{\ga}_u^\prime=\ga_1\sin^2\beta, &\qquad  \tilde{\ga}_d=\ga_1\cos^2\beta\\
h^u=\lambda^{u*}\sin\beta, &\qquad \hspace{-.2cm}
h^{d,e}=\lambda^{d,e*}\cos\beta
\end{split}
\end{equation}
where  $\tan\beta$ is the ratio of the two vev's.

In solving the  two-loop RGE's for the gauge
couplings ($\ga_i, i=1,2,3$),  gaugino couplings
($\ga_u,\ga_d,\ga_u^\prime,\ga_d^\prime$) and the Yukawa couplings
$\lambda^u$ and $h^u$, we followed the bottom-up approach
beginning at the scale $M_Z$ where the initial conditions for the gauge and the Yukawa
couplings are well known experimentally\footnote{We have taken into account the running of the Yukawa coupling from $m_{\rm{top}}$ to $M_Z$}.
The missing initial conditions for
the gaugino couplings at $M_Z$ can be determined as follows. First, we construct an exact power series solution for the relevant RGE's. Then,
we approximate the truncated power series by means of Pade approximants. The sought initial conditions can be found by solving (\ref{matching}).
Finally, we numerically (re)solved  the whole system requiring the
matching at $\tilde{m}$ to be valid within an error of less than 1\%. In the actual running the error was less than 0.4 \% .

\begin{figure}[!t]
\begin{tabular}{@{}m{7cm}m{7cm}}
\includegraphics[scale=.85]{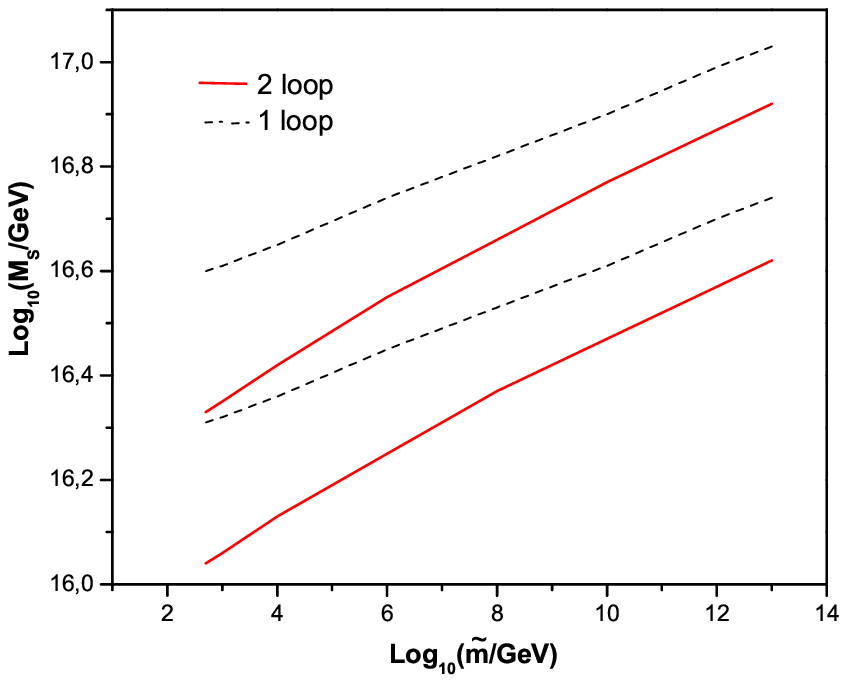}&
\includegraphics[scale=.85]{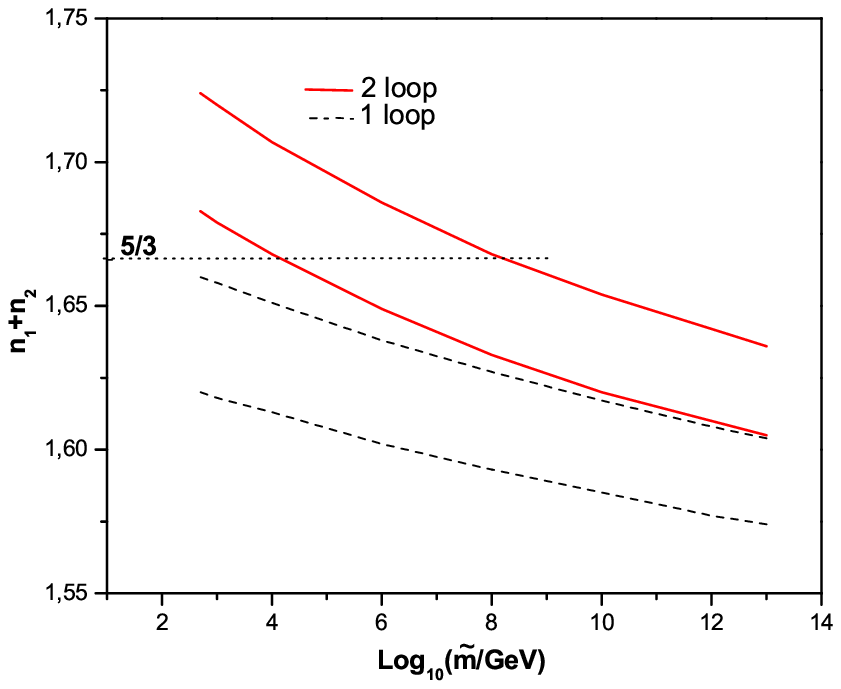}\\
\hspace*{4cm}(a)&\hspace*{4cm}(b)
\end{tabular}
\caption{The $M_{S}$ scale,(a), and  the sum $n_1+n_2$, (b),
as a function of $\log_{10}\tilde{m}$ for the case where
$\alpha_3=\alpha_2$ at $M_{S}$. The bands correspond to the
strong coupling experimental error at $M_Z$. The solid lines
correspond to 2-loop running while the dotted ones to 1-loop.}
\label{Graph_1}
\end{figure}

We begin by first considering  the case of
$SU(3)$ and $SU(2)$ gauge couplings unification i.e., when
$\xi=1$.  In this case, $k_Y=n_1+n_2=\frac{\alpha_s}{\alpha_1}$  where
$\alpha_s$ is the unified gauge coupling at $M_S$. It can be checked that
several models in Table (\ref{t1}) predict the standard $SU(5)$
normalization  $k_Y=5/3$.  In Fig.(\ref{Graph_1}) we plot $M_{S}$, in (a) and the sum
$n_1+n_2$, in (b), as functions of
 $\tilde{m}$. The band delimits the $\alpha_3(M_Z)$
experimental uncertainty. The solid lines correspond to the 2-loop running
while the dotted ones to the 1-loop running. In (b) we also show the
line $n_1+n_2=5/3$. This
figure confirms that the condition $a_3=a_2$  is satisfied for
$M_S\sim 10^{16}$GeV, irrespectively of the number of $U(1)$ branes
added in the configuration, provided that the specific setup ensures
that $k_Y=\frac 53$.

The  $M_{S}$ scale found here is expected to be higher
than the corresponding unification scale of the MSSM. This is due to the presence,
in the split case and below $\tilde{m}$, of extra degrees of freedom (gauginos and higgsinos)
which modify the $\beta$ functions.
The width of the band which corresponds to the strong coupling experimental error at
$M_Z$ corresponds almost to a factor of 2 in the $M_S$ scale. This is not at all surprising,
since simple calculations in 1-loop approximation show that
\[
\ln M_S=\ln \tilde{m}+\left(\frac{1}{\alpha_2(M_Z)}-\frac{1}{\alpha_3(M_Z)}\right)
                 \frac{2\pi}{b_2^S-b_3^S}-
                 \frac{b_2^{SP}-b_3^{SP}}{b_2^S-b_3^S}\left(\ln\tilde{m}-\ln M_Z\right)
\]
where the superscript S (SP)to the b-coefficients corresponds to the Susy (Split Susy) case.
The difference in $\ln M_S$, due to the strong coupling error and for constant
m is given by
\[
\frac{\Delta^{strong}M_S}{M_S}=
\Delta^{strong}\left(\frac{1}{\alpha_2(M_Z)}-\frac{1}{\alpha_3(M_Z)}\right) \frac{2\pi}{b_2^S-b_3^S}=
\sim 0.666
\]

\begin{figure}[!b]
\centering
\includegraphics[scale=.85]{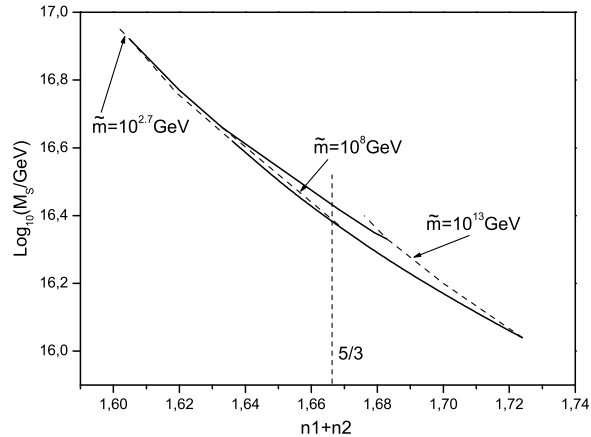}
\caption{The $M_S$ scale as a function of $n_1+n_2$. The two curves correspond to the $a_3(M_Z)$ experimental
uncertainty. Curves of constant $\tilde{m}$ are shown for the values $10^{2.7}$,
$10^8$ and $10^{13}$ GeV.}
\label{Graph_2}
\end{figure}

The correlation of the string and Split Susy
scales for different values of $k_Y$, can be seen in
Fig.(\ref{Graph_2}) where, we plot $M_S$ as a function of $k_Y$.  It can be readily seen that the  $M_{S}$ value
is not very sensitive to the value of $k_Y$.

\begin{figure}[!t]
\begin{tabular}{@{}m{7cm}m{7cm}}
\includegraphics[scale=0.8]{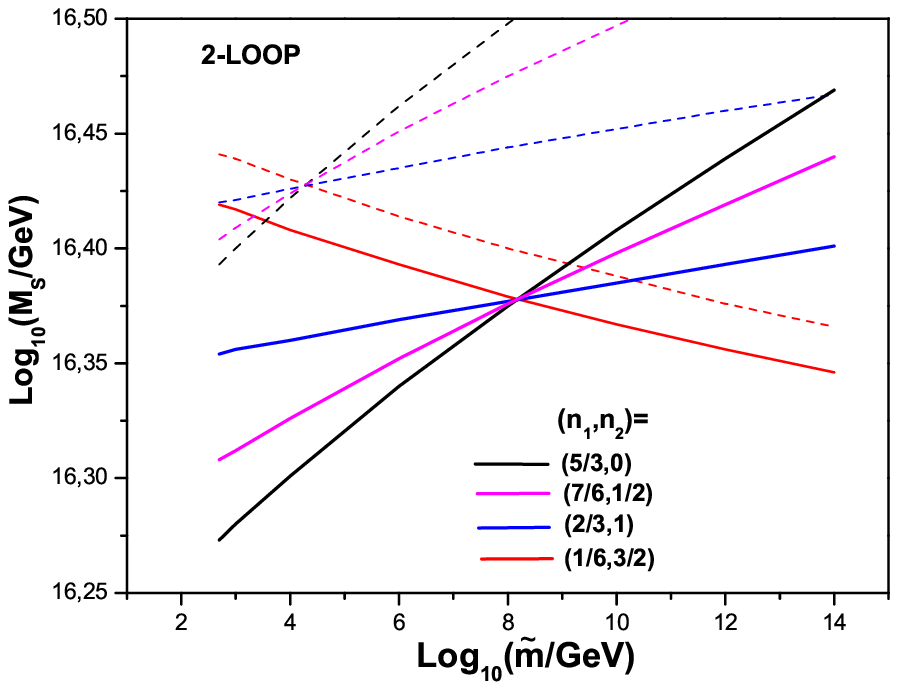}&
\includegraphics[scale=0.8]{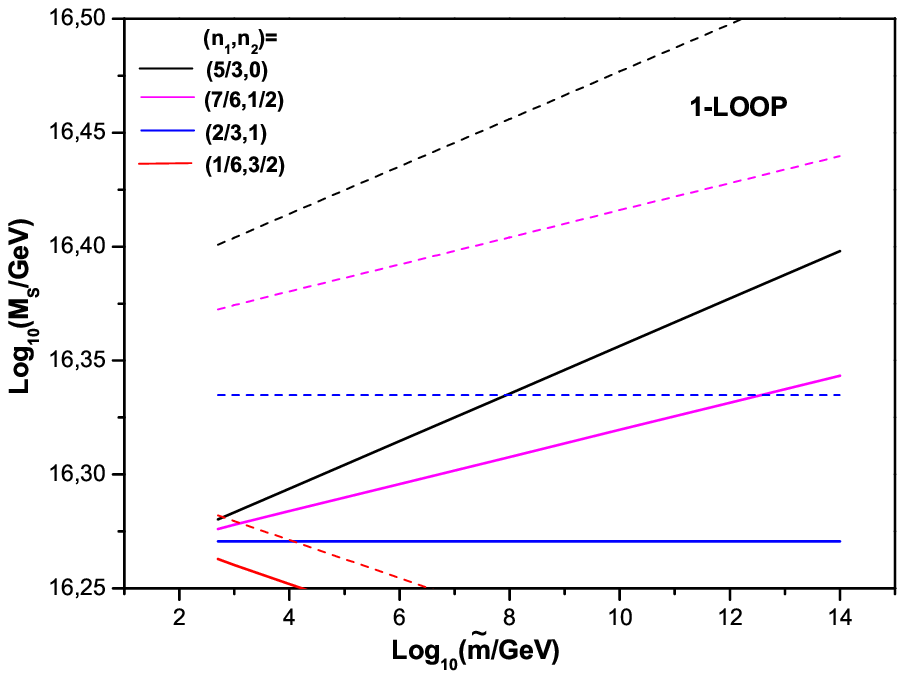}\\
\hspace*{4cm}(a)&\hspace*{4cm}(b)
\end{tabular}
\caption{The $M_{S}$ scale as a function of $\tilde{m}$, for
2-loop running (a) and 1-loop running (b), for the models having
$n_1+n_2=5/3$. Continuous lines correspond to the lower value of
$\alpha_3(M_Z)$ while dotted ones to the higher one.}
\label{Graph_3}
\end{figure}

Next,  relaxing the requirement of $SU(3)$ and $SU(2)$ gauge
coupling unification, we calculate  $M_{S}$ for all
models that satisfy the relation $n_1+n_2=5/3$ as a function of $\tilde{m}$.
Since the equality of $\alpha_2$ and $\alpha_3$ at
$M_{S}$ is no longer required, different models evolve differently.
The results are depicted in Fig.(\ref{Graph_3}). It is interesting to realize that
all the curves have one common point. In
other words, there exists a singled out $\tilde{m}$ value that gives the same
$M_{string}$ for all models in this class. This is expected as we
have seen in our first approach and in Fig.(\ref{Graph_2}): If we
choose a specific $\tilde{m}$ scale we can satisfy
Eq.(\ref{atstring}) at a scale $M_{S}$ with the additional
requirement that at this  scale $\alpha_3=\alpha_2$. All
models corresponding to a given $n_1+n_2$ behave similarly.
This behavior is independent of 1-loop or 2-loop running, although
these special points are different (see Fig.(\ref{Graph_3}(b)).
Notice also the big difference (6 orders of magnitude) in the value of
$\tilde{m}$ between the corresponding meeting point of 1- and 2-loop
running although the difference in $M_{S}$ is much smaller.
This point is also made clear in Fig.(\ref{Graph_1}b), where the curves
meet the $5/3$ value at very low $\tilde{m}$.

\begin{figure}[!t]
\centering
\includegraphics[scale=0.85]{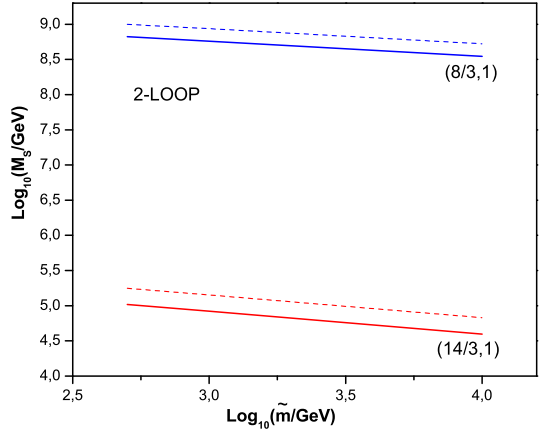}
\caption{The $M_{S}$ scale as a function of $\tilde{m}$, for
2-loop running for the models with (8/3,1) and (14/3,1). Continuous
lines correspond to the lower value of $\alpha_3(M_Z)$ while dotted
ones to the higher one.} \label{Graph_4}
\end{figure}

In Fig.(\ref{Graph_4}) we  show for the sake of completeness, the corresponding graph for
two more models, namely (8/3,1) and (14/3,1). The values for both
$M_{S}$ and $\tilde{m}$ are now found to be much smaller than before.

\section{Mirrors}

\begin{table}[!b]
\centering
\[
\begin{array}{|c|c|c|c|c|c|c|c|c|}
          \hline
\textrm{model}
  &          Q & U & D & L & E &  N  &   Y & H \\
          \hline
1 &          0 & 0 & 0 & 0 & 2 &  3  &   2 & 16\\
          \hline
2 &          0 & 0 & 0 & 0 & 2 &  3  &   6 & 16\\
          \hline
3 &          0 & 0 & 0 & 2 & 2 &  3  &   6 & 8\\
          \hline
4 &          0 & 0 & 0 & 2 & 2 &  3  &   6 & 24\\
          \hline
5 &          0 & 0 & 0 & 6 & 2 &  3  &   2 & 8\\
          \hline
6 &          0 & 0 & 0 & 6 & 2 &  3  &   2 & 24\\
          \hline
7 &          2 & 0 & 0 & 0 & 2 &  3  &   2 & 16\\
          \hline
8 &          2 & 0 & 0 & 0 & 2 &  3  &   6 & 16\\
          \hline
9 &          4 & 0 & 0 & 2 & 2 &  3  &   2 & 8\\
          \hline
10 &          4 & 0 & 0 & 2 & 2 & 3  &   2 & 24\\
          \hline
11 &          4 & 0 & 0 & 6 & 2 & 3  &   6 & 8\\
          \hline
12 &          4 & 0 & 0 & 6 & 2 & 3  &   6 & 24\\
          \hline
\end{array}
\]
\caption{The twelve possible configurations of extra mirror content}
\label{T1}
\end{table}

As mentioned previously, D-brane constructions include states with the quantum numbers of
leptoquarks. Such cases have already been analyzed in the context of CFT-orientifolds models.
Here, we will pick up the 32 models presented in
\cite{Ibanez:2007rs}
which also have the advantage of inducing neutrino Majorana masses through instanton effects.
The extra particles of these models consist of mirror states
having the same quantum numbers (under the SM gauge group) as the standard ones, with the exception
of singlet neutrinos and d-like leptoquarks, i.e. states with the quantum numbers of d-quarks carrying both lepton and
baryon number. Four brane-stack configurations are considered in the above construction, two of them carrying the $SU(3)$ and the
$SU(2)$, with 2 more providing either a $U(1)$ and an $O(2)$, or two $U(1)$ factors.
A similar analysis can also be carried out for the five brane-stack scenario discussed in the present work.

The above 32 models correspond to only 12 distinct spectra with the exotic states shown in Table \ref{T1}.
The doublet and the singlet quarks are denoted by $Q$, $U$ and $D$ respectively, the doublet and the singlet leptons
are  $L$ and $E$, $N$ are the right-handed neutrinos,  $Y$ stands for the leptoquarks and finally, $H$ represents the Higgs.
Note that in these models the minimum  number of RH neutrinos required is 3 (to avoid the cubic anomaly) however, since
these are singlets under the SM gauge group do not contribute to the running of the couplings.

We will now repeat the renormalization group analysis for these models, including also the exotic states.
Starting again from the known initial values of the three SM couplings at $M_Z$, we determine, for each
of the above models, the pair $(\tilde{m},M_{S})$ subject to the ``unification'' condition (\ref{atstring}) at $M_{S}$.
Above  $\tilde{m}$ we have the complete MSSM spectrum enriched by the extra mirrors appearing
in Table \ref{T1}. Below $\tilde{m} $, we have  Split Susy containing   extra
quarks, leptons, leptoquarks and higgsinos, while we keep one Higgs doublet only as in the original Split scenario. Since our group is
the SM one, no extra gauginos appear.

\begin{figure}[!t]
\begin{tabular}{@{}m{7cm}m{7cm}}
\includegraphics[scale=0.85]{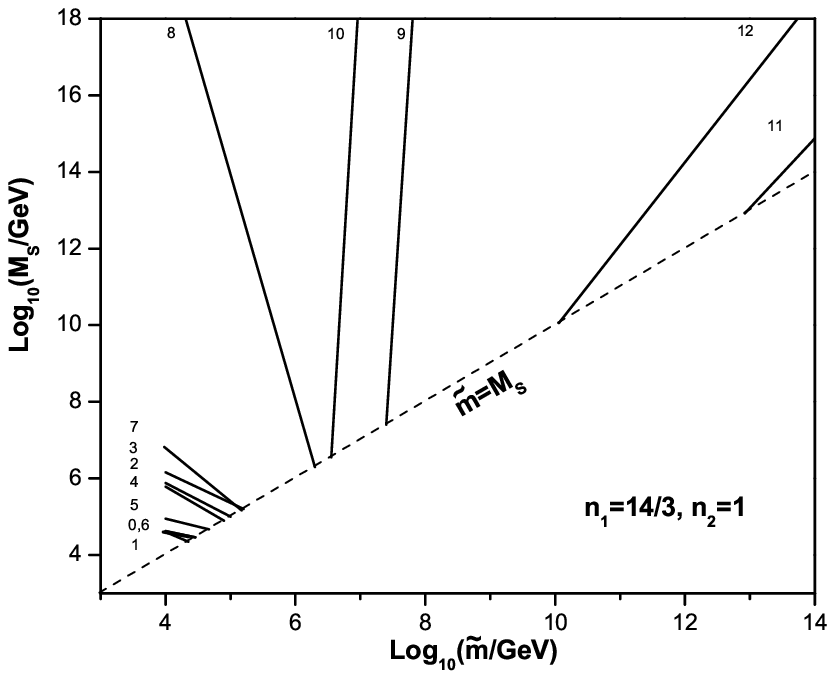}&
\includegraphics[scale=0.85]{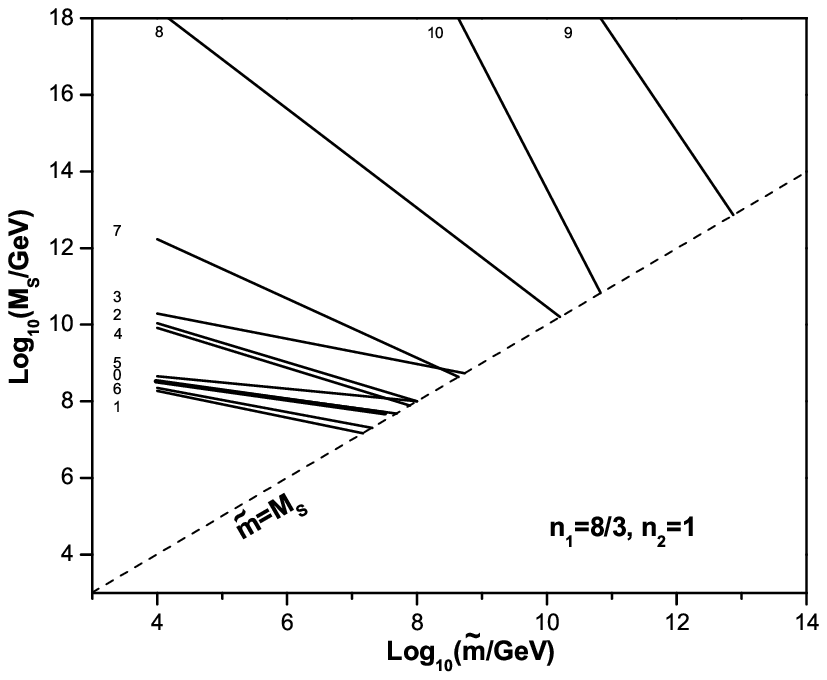}\\
\hspace*{4cm}a & \hspace*{4cm}b
\end{tabular}
\caption{Plots of $M_{S}$ vs $\tilde{m}$ for the twelve models appearing in Table \ref{T1}
and the two cases of the pair $(n_1,n_2)$. The thicker line corresponds to the case with no mirrors
(0 model) while the label corresponds to the model as in Table \ref{T1}.}
\label{mirrors_1}
\end{figure}

In Fig.\ref{mirrors_1} we plot the one-loop results for the 12
models together with the case of no extra mirrors (thick line)
and for $(n_1,n_2)$ equal to(14/3,1) and (8/3,1) respectively.
The allowed region lies above the $\tilde{m}=M_{S}$ line.
The emerging picture seems to favor  higher $M_{S}$ values as compared to the minimal cases
discussed previously.
Notice that in models 9 and 10, $\tilde m$ is almost fixed ($10^{6.5}$ and $10^{7.5}$)
while $M_{string}$ varies along the whole (acceptable) range.

Some comments are here in order. In Fig.\ref{mirrors_1}(a), the depicted models can be
grouped in three distinctive classes: (i) In the low $M_S$ class, the higher $\tilde m$ the lower
$M_S$. (ii) In the class with high $M_S$ the tendency is reversed. (iii) The third class
has almost constant $\tilde m$ while $M_S$ varies along the whole (acceptable) range.
In Fig.\ref{mirrors_1}(b), the majority of the models show that $M_S$ is less sensitive against the variation
of $\tilde m$. This rather peculiar behavior can be easily explained if we recall Eq.(\ref{stringscale}) written in the form
\[
\ln \left[ {\frac{{M_S }}{{M_Z }}} \right] = a\ln \left[ {\frac{{\tilde m}}{{M_Z }}} \right] + b \, \cdot
 \]
In models of the first class $a<0$. In class (ii) $a>0$ and $M_S$ increases with ${\tilde m}$.
Finally, for models in (iii) $a $ is very large and $M_S$ is almost independent of  ${\tilde m}$.

\section{The lepton flavor violating process  $\mu\rightarrow
e\gamma$}

One of the most important flavor violating rare processes
 present in all models with flavor mixing in the Yukawa
sector, is the  decay of the muon to an electron-photon pair.
In ordinary supersymmetric theories, this exotic
process is usually suppressed by powers of the supersymmetry
breaking scale which is of the order of 1 TeV at the most
\cite{Hisano:1995cp},\cite{Ellis:1999uq}. If the scalar partners
involved in the process are non-universal, hard violations of
the lepton number conservation that exceed the present experimental
limits are expected.  Even in
the case of universal masses at the unification scale,
renormalization effects result to mass splitting and important
mixing effects  at low energies. Although there is a certain degree
of  ambiguity due to the unknown mixing details in lepton and
slepton mass matrices in the calculation of the branching ratios, a
significant portion of the
universal gaugino mass--universal scalar mass
parameter space $(m_{1/2},m_0)$ is
excluded in ordinary supersymmetric models. In the present analysis,
we examine the conditions for accessibility of this interesting
decay in the context of the D-brane constructions discussed above.

In Split Susy, the scale of the slepton masses is set by the $\tilde{m}$
 scale which is considered to be much higher than the TeV ordinary
Susy scale. The renormalization group analysis of the generic
D-brane constructions has shown however, that in several cases the
$\tilde{m}$ can be as low as a few TeV. Graphs for
$\mu\rightarrow e\gamma$ are mediated by the above scalars.
It is worth exploring whether in some of the above models, these exotic
reactions could be observed in future experiments.
The branching ratio of the $\mu\rightarrow e\gamma$ process
is given by \cite{Hisano:1995cp,Ellis:1999uq}
\[
\textrm{BR}(\mu\rightarrow
e\gamma)=\frac{48\pi^3\alpha}{G_F^2}\left(|A_2^L|^2+|A_2^R|^2\right)
\]
where
\[
A_2^L=A_2^{(n)L}+A_2^{(c)L}\quad\quad \textrm{and}\quad\quad
A_2^R=A_2^{(n)R}+A_2^{(c)R}\, \cdot
\]
The superscript $n$ ($c$) of the amplitudes $A$ corresponds to the
neutralino (chargino) exchange  contribution and $L$ ($R$) to the
left (right) handed incoming lepton. The amplitudes are given by the
relations
\begin{align}
A_2^{(n)L}&= \frac{1}{32\pi^2}\frac{1}{\tilde{m}^2_{\tilde{l}_X}}
\left[ N^{L(l)}_{1AX} N^{L(l)*}_{2AX}\,f_1(r) + N^{L(l)}_{1AX}
N^{R(l)*}_{2AX}\,\frac{M_{\tilde{\chi}^0_A}}{m_\mu}\,f_2(r)
\right]\label{An}\\
A_2^{(c)L}&= -\frac{1}{32\pi^2}\frac{1}{\tilde{m}^2_{\tilde{\nu}_X}}
\left[ C^{L(l)}_{1AX} C^{L(l)*}_{2AX}\,f_3(r') + C^{L(l)}_{1AX}
C^{R(l)*}_{2AX}\,\frac{M_{\tilde{\chi}^-_A}}{m_\mu}\,f_4(r')
\right]\label{Ac}\\
A_2^{(n)R}&=A_2^{(n)L}|_{L\leftrightarrow
R}\quad\quad\textrm{and}\quad\quad
A_2^{(c)R}=A_2^{(c)L}|_{L\leftrightarrow R}\, \cdot
\end{align}
We are working with the mass eigenstates so, all matrices rotating
from the weak to the mass eigenstates are moved to the vertices. The neutral
amplitude involves the neutralinos and the charged sleptons with
corresponding masses $M_{\tilde{\chi}^0_A}$ ($A=1,...,4$) and
$\tilde{m}_{\tilde{l}_X}$ ($X=1,...,6$) while $m_\mu$ is the mass of
the decaying muon. The charged amplitude involves the charginos and
sneutrinos with corresponding masses $M_{\tilde{\chi}^-_A}$
($A=1,2$) and $\tilde{m}_{\tilde{\nu}_X}$ ($X=1,2,3$). The matrices
$N^{L(l)}_{iAX}$ and $N^{R(l)}_{iAX}$ involve both the neutralino
and the slepton rotating matrices ($i=1$ for the electron and $i=2$
for the muon) and are given by
\[
\begin{split}
N^{R(l)}_{iAX}&=-\frac{g_2}{\sqrt{2}} \left\{ \left[
-(O_N)_{A2}-(O_N)_{A1}\tan\theta_W\right]U^l_{Xi}+
\frac{m_i}{m_W\cos\beta}\,(O_N)_{A3}U^l_{X(i+3)}\right\}\\
N^{L(l)}_{iAX}&=-\frac{g_2}{\sqrt{2}} \left\{
\frac{m_i}{m_W\cos\beta}\,(O_N)_{A3}U^l_{Xi}+
2(O_N)_{A1}\tan\theta_WU^l_{X(i+3)}\right\} \, \cdot
\end{split}
\]
The matrices $O_N$ rotate the neutralinos while $U^l$ rotate the
charged sleptons. Finally, $m_i$ is the lepton mass ($m_1=m_e$,
$m_2=m_\mu$).

The matrices $C^{L(l)}_{iAX}$ and $C^{L(l)}_{iAX}$ (again $i=1$ for
the electron and $i=2$ for the muon) involve both the chargino
($O_L$ and $O_R$) and the sneutrino ($U^\nu$) rotating matrices and
they are given by
\[
\begin{split}
C^{R(l)}_{iAX}&=-g_2(O_R)_{A1} U^\nu_{Xi}\\
\\
C^{L(l)}_{iAX}&=g_2\,\frac{m_i}{\sqrt{2}m_W\cos\beta}(O_L)_{A2}U^\nu_{Xi}\, \cdot
\end{split}
\]
Finally, the $f$ functions appearing in Eqs.(\ref{An}) and
(\ref{Ac}) are
\[
\begin{split}
f_1(r)&=\frac{1}{6(1-r)^4}(1-6r+3r^2+2r^3-6r^2\ln(r))\\
f_2(r)&=\frac{1}{(1-r)^3}(1-r^2+2r\ln(r))\\
f_3(r')&=\frac{1}{6(1-r')^4}(2+3r'-6r'^2+r'^3+6r'\ln(r'))\\
f_4(r')&=\frac{1}{(1-r)^3}(-3+4r'-r'^2-2\ln(r'))\\
\textrm{where,}\quad
r&=\frac{M^2_{\tilde{\chi}^0_A}}{\tilde{m}^2_{\tilde{l}_X}}\quad\quad\textrm{and}\quad
\quad r'=\frac{M^2_{\tilde{\chi}^-_A}}{\tilde{m}^2_{\tilde{\nu}_X}}\, \cdot
\end{split}
\]

In estimating the $\mu\rightarrow e\gamma$ branching ratio
and in order not to complicate our analysis we have ignored the neutralino and the
chargino mixing\footnote{
Actually, an explicit calculation for the models under consideration, shows that inclusion
of the neutralino and chargino mixing matrices modifies the
branching ratio by less that 10\%.}
(i.e. the $O_L$, $O_R$ and $O_N$ matrices are the
identity matrix) and have dropped terms proportional to the mass of the
leptons. Therefore, we are left with the $U^l$ and $U^\nu$ matrices.

The $6\times 6$ charged slepton matrix is given by
\begin{equation}
\left(
\begin{array}{cc}
m^2_{LL}  & m^2_{LR}\\
m^2_{RL}  & m^2_{RR}\\
\end{array}
\right)
\end{equation}
where each entry is a $3\times 3$ matrix with an obvious notation and \cite{Gomez:1998wj}
\begin{align}
m^2_{LL}&=(\tilde{m}^\delta_{\tilde{l}})^2+\delta m^2_N+m^2_l+M^2_Z\left(\frac12-\sin^2\theta_W\right)\cos 2\beta\\
m^2_{RR}&=(\tilde{m}^\delta_{\tilde{e}_R})^2+m^2_l-M^2_Z\sin^2\theta_W\cos 2\beta\\
m^2_{RL}&=(A_e^\delta+\delta A_e+\mu\tan\beta)m_l\\
m^2_{LR}&=m^{2\dag}_{RL} \, \cdot
\end{align}
The superscript $\delta$ denotes the diagonal part of the
corresponding $3\times 3$ matrix and we have assumed universal soft masses,
so that
$(\tilde{m}^\delta_{\tilde{l}})^2=(\tilde{m}^\delta_{\tilde{e}_R})^2=
\textrm{Diagonal}(m_0,m_0,m_0)$ where in our case, $m_0=\tilde{m}$.
Also, $A_e^\delta=\textrm{Diagonal}(A_0,A_0,A_0))$ with $A_0=-1.5m_0$.
The terms $\delta m^2_N$ and $\delta A_e$
stand for the off-diagonal matrices which arise because of  the
non-diagonal Yukawa coupling $\lambda_D$ (in the basis where the
lepton matrix is diagonal) appearing in the term which gives mass to
the neutrino through the superpotential term $N^c\lambda_D lH_2$ and
in the trilinear coupling of the potential correspondingly. The
sneutrino reduced $3\times 3$ matrix (\cite{Gomez:1998wj}) is given by
\begin{equation}
\tilde{m}^2_{\tilde{\nu}}=(\tilde{m}^\delta_{\tilde{l}})^2+\delta
m^2_N+\frac12M^2_Z\cos 2\beta \, \cdot
\end{equation}

\begin{figure}[!t]
\centering
\includegraphics[scale=0.85]{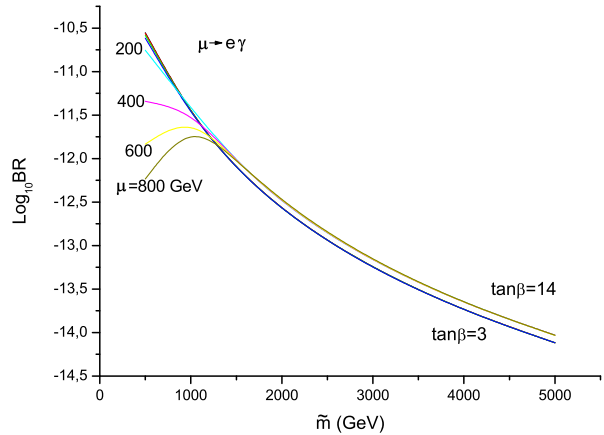}
\caption{The (logarithm of the) branching ratio of $\mu\rightarrow
e\gamma$ as a function of $\tilde{m}$ for $\tan\beta=3$ and 14 and
for $\mu=200,400,600,800$ GeV.} \label{Graph_5}
\end{figure}

The off diagonal matrices $\delta m^2_N$ and $\delta A_e$ are
evaluated in \cite{Gomez:1998wj} for the region of $\tan\beta=3-14$.
Having determined the  mass matrices for the charged sleptons and
the sneutrino, we can find the matrices $U^l$ and $U^\nu$ which
rotate to the corresponding mass eigenstates and therefore, proceed
to the evaluation of the desired branching ratio. In
Fig.(\ref{Graph_5}) we show the BR as a function of $\tilde{m}$ for
the above region of $\tan\beta$ and for $\mu=200-800$ GeV. We
clearly see that for the present experimental BR bound ($10^{-12}$),
$\mu\rightarrow e\gamma$ could be observed in models allowing
$\tilde m$ around 1.5 TeV
(for the evaluation of the BR we have assumed a universal gaugino mass
of 200 GeV, since we run our RGE's with the Split Susy regime active down to the weak scale).

From the renormalization group analysis of the previous section, we
can check that for a wide class of D-brane constructions, the
Split-supersymmetric scale ranges from 1 TeV up to values as high as
$10^{13}$ GeV.  The above reaction could in principle be observed in
the present day experiments for the lower marginal
values of the Split Susy scale as it can also be seen  from Fig. \ref{Graph_1}
and \ref{Graph_2}. However, if we stick to  cases where we have $a_3=a_2$ unification at
$M_S$, the $U(1)_Y$ normalization constant $k_Y$ is found to be around the
standard value $\frac 53$. Then, from Fig. \ref{Graph_2} we  observe that
$\tilde m\sim 10^{8}$ GeV therefore, the branching ratio is significantly
suppressed and lies far beyond the capabilities of even  future experiments.

On the other hand, the observability prospects of this reaction are different in models
implying low string scale, since
the Split scale $\tilde m$ cannot be higher that $\tilde m\le M_S\sim 5$ TeV.
For $\tilde m\sim 5$ TeV,  the ${\mu\ra e\gamma}$ branching ratio is
$BR_{\mu\ra e\gamma}\sim 10^{-14}$, a number that could in principle be checked
in future experiments. Thus, for this particular class of D-brane constructions,
muon number violating  reactions  can probe  the
whole $\tilde m$-region.

\section{Conclusions}
D-branes brought profound changes in our approach to model building.
In the present work, we presented the simplest D-brane extensions of
the Standard Model based on  configurations with gauge symmetry of
the form $U(3)\times U(2)\times U(1)^P$  with $P\le 3$. Exploring
the different  brane-stack orientations in the higher dimensional
space as well as the various hypercharge embeddings consistent with
the SM particle spectrum, we managed to obtain a complete
classification of the SM variants that arise in this context. We
addressed several phenomenological issues affecting the viability of
these constructions. These issues were examined in the context of
split supersymmetric spectrum which has been shown to be a  natural
possibility in a wide class of D-brane constructions. This way, we
extended previous investigations on the correlation between the
string scale and the split supersymmetry breaking scale,
incorporating two-loop effects. The calculations show a significant
change on the split supersymmetry breaking scale as compared to the
one-loop results, while, the string scale for several interesting
cases is only moderately affected. The analysis was further extended
to non-minimal cases that include leptoquark and mirror states.
Yukawa terms providing masses to the SM fields were  calculated,
while  the problem of incorporating  the right handed neutrino mass through
instanton effects was also mentioned. Finally, we presented a
 detailed discussion on the possible observability of  the $\mu\ra e\gamma$
flavor violating decay in the present and future experiments.

\vspace*{0.5cm}
\noindent
The work is partially supported by the EU Grant MRTN-CT-2004-503369.

\end{document}